\documentclass{emulateapj}

\usepackage{amsmath}
\usepackage[]{natbib}
 
\newcommand{\kms}{\>{\rm km}\,{\rm s}^{-1}}

\newcommand{\Msun}{\mbox{$\rm M_{\odot}$}}
\newcommand{\Mbh}{\mbox{$\rm M_{\bullet}$}}
\newcommand{\Msig}{\mbox{$\rm M_{\bullet}-\sigma$}}

\newcommand\degrees{^\circ}

\newcommand{\etal}{{et al.~}}

\newcommand\ie{{\it i.e.}}

\newcommand\hst{{\it HST}}

\slugcomment{{\it Unfinished draft} \today}

\shorttitle{The Binary Nucleus in VCC 128}  

\shortauthors{Debattista \etal}

\begin{document}

\title{The Binary Nucleus in VCC 128: A Candidate Supermassive
Black Hole in a Dwarf Elliptical Galaxy}

\author{Victor P. Debattista,\altaffilmark{1,2}
Ignacio Ferreras,\altaffilmark{3}
Anna Pasquali,\altaffilmark{4}
Anil Seth,\altaffilmark{1,5}
Sven De Rijcke,\altaffilmark{6} and
Lorenzo Morelli\altaffilmark{7}} 

\altaffiltext{1}{Astronomy Department, University of Washington, Box
351580, Seattle WA 98195, USA {\tt debattis@astro.washington.edu}}

\altaffiltext{2}{Brooks Prize Fellow}

\altaffiltext{3}{Physics Department, King's College London, Strand,
London, WC2R 2LS, UK {\tt ferreras@star.ucl.ac.uk}}

\altaffiltext{4}{Max-Planck-Institut f\"ur Astronomie, Koenigstuhl 17,
D-69117 Heidelberg, Germany {\tt pasquali@mpia-hd.mpg.de}}

\altaffiltext{5}{Currently a CfA Postdoctoral Fellow, 60 Garden St.,
Cambridge, MA 02138 {\tt aseth@cfa.harvard.edu}}

\altaffiltext{6}{Sterrenkundig Observatorium, Universiteit Gent,
Krijgslaan 281, S9, B-9000 Gent, Belgium {\tt Sven.DeRijcke@UGent.be}}

\altaffiltext{7}{Departamento de Astronomia y Astrofisica, Pontificia
Universidad Catolica de Chile, Casilla 306, Santiago 22, Chile {\tt
lmorelli@astro.puc.cl}}

\begin{abstract}
{\it Hubble Space Telescope} (\hst) Wide Field Planetary Camera 2
({\it WFPC2}) images of the Virgo Cluster dwarf elliptical galaxy VCC
128 reveal an apparently double nucleus.  The two components, which
are separated by $\sim 32$ pc in projection, have the same magnitude
and color.  We present a spectrum of this double nucleus and show that
it is inconsistent with one or both components being emission-line
background objects or foreground stars.  The most likely
interpretation is that, as suggested by \citet{lauer_etal_96} for the
double nucleus in NGC 4486B, we are seeing a nuclear disk surrounding
a supermassive black hole.  This is only the second time an early-type
dwarf (dE/dSph) galaxy has been suggested to host a SMBH.
\end{abstract}


\section{Introduction}
\label{s:intro}

The centers of galaxies often host supermassive black holes (SMBHs)
with masses \Mbh\ ranging from $\sim 10^6$ to $\sim 10^9$ \Msun\
\citep{kor_ric_95}.  SMBH masses exhibit a variety of scaling
relations including the \Msig\ relation \citep{geb_etal_00,
fer_mer_00} and an $\Mbh-M_{DM}$ relation, where $M_{DM}$ is the mass
of the host dark halo \citep{ferrar_02, bae_etal_03}.  These
correlations are evidence for a coupling between the formation of
SMBHs and galaxies.  One key to unraveling this coupling is to
identify the seeds of SMBHs in dwarf galaxies.  Ground-based stellar
kinematic data of dwarf elliptical galaxies (dEs) \citep{geh_etal_02}
rule out $\Mbh \ga 10^7$ \Msun.  In the M31 globular cluster G1,
thought to be the stripped nucleus of an accreted dE,
\citet{geb_etal_02} report finding an intermediate mass black hole
(IMBH) with $\Mbh \sim 10^4~\Msun$, consistent with the extrapolation
of the \Msig\ relation.  Nuclear activity in low-luminosity dwarf
galaxies has turned up a number of IMBHs with estimated masses from
$10^4$ to $10^6~ \Msun$ \citep{fil_ho_03, bar_etal_04, gre_ho_04}.
There is instead no evidence of a SMBH in either M33
\citep{mer_etal_01, geb_etal_01} or NGC 205 \citep{val_etal_05}.
The paucity of SMBHs in dwarf galaxies led \citet{ferrar_02} to
suggest that low mass galaxies are inefficient at forming SMBHs.
\citet{cot_etal_acsvcs8_06}, \citet{fer_etal_06} and
\citet{weh_har_06} find instead a continuity in scaling relations,
with \Mbh\ replaced by $M_{\rm CMO}$, the mass of a central massive
object, which is either a SMBH or a compact nucleus.

\hst\ does not permit many dynamical \Mbh\ measurements below $\sim
10^6~ \Msun$.  An alternative approach to finding SMBHs in low surface
brightness dwarf galaxies is a morphological one.  In a few galaxies,
\hst\ has found double nuclei; the clearest examples are in M31 and in
NGC 4486B \citep{lauer_etal_93, lauer_etal_96, lauer_etal_05}.  The
two nuclei in M31, dubbed P1 and P2, are separated by $0\farcs 49$ or
3.6 pc and have unequal surface brightness, with P2 having the lower
surface brightness but sitting very close to the global photocenter.
In contrast, in NGC 4486B the two nuclei are equal in magnitude, color
and displacement from the photocenter, with a total separation of 12
pc.
The generally accepted model of the nucleus in M31 identifies P2 as
the center of the galaxy surrounding its SMBH and P1 as the bright
off-center apoapsis of an eccentric Keplerian disk \citep{tremai_95}.
Self-consistent models and simulations of such nuclei can be
constructed \citep{sal_sta_01, jac_sel_01, sam_sri_02}.
\citet{lauer_etal_96} proposed a similar model for the double nucleus
in NGC 4486B with the difference that the SMBH is located on the
global photocenter between the two peaks.  They suggested that binary
nuclei are morphological indicators of SMBHs.

Here we present another example of a galaxy with a double nucleus, the
Virgo Cluster dE VCC 128.  Its nucleus is very similar to that of NGC
4486B.  Section \ref{sec:hst} presents the archival \hst\ data,
Section \ref{sec:sed} presents new spectra and their modeling and in
Section \ref{sec:discussion} we propose that the double nucleus is
best explained by a disk surrounding a SMBH at the center of VCC 128
and explore its likely position in the \Msig\ plane.

\section{Analysis of \hst\ Archival Images}
\label{sec:hst}

Using the \hst\ archive, we explored the nuclear morphologies of the
sample of dEs imaged in the {\it WFPC2} snapshot surveys Prop-ID
GO-8600 and GO-6352 (PI Ferguson).  These surveys imaged 50 galaxies,
which were analysed by \citet{lot_etal_04} to show that typical dE
stellar envelopes are 0.1-0.2 mag. redder in $V-I$ than their nuclei.
In most of the galaxies we either found no nucleus or a single
nucleus.  VCC 1107, FCC 208 and VCC 128 contained what appeared to be
double nuclei; of these the two nuclei in VCC 128 are the closest, and
are most similar in magnitudes making this galaxy a prime candidate
for spectroscopic follow-up.

\begin{figure*}[!htbp]
\plotone{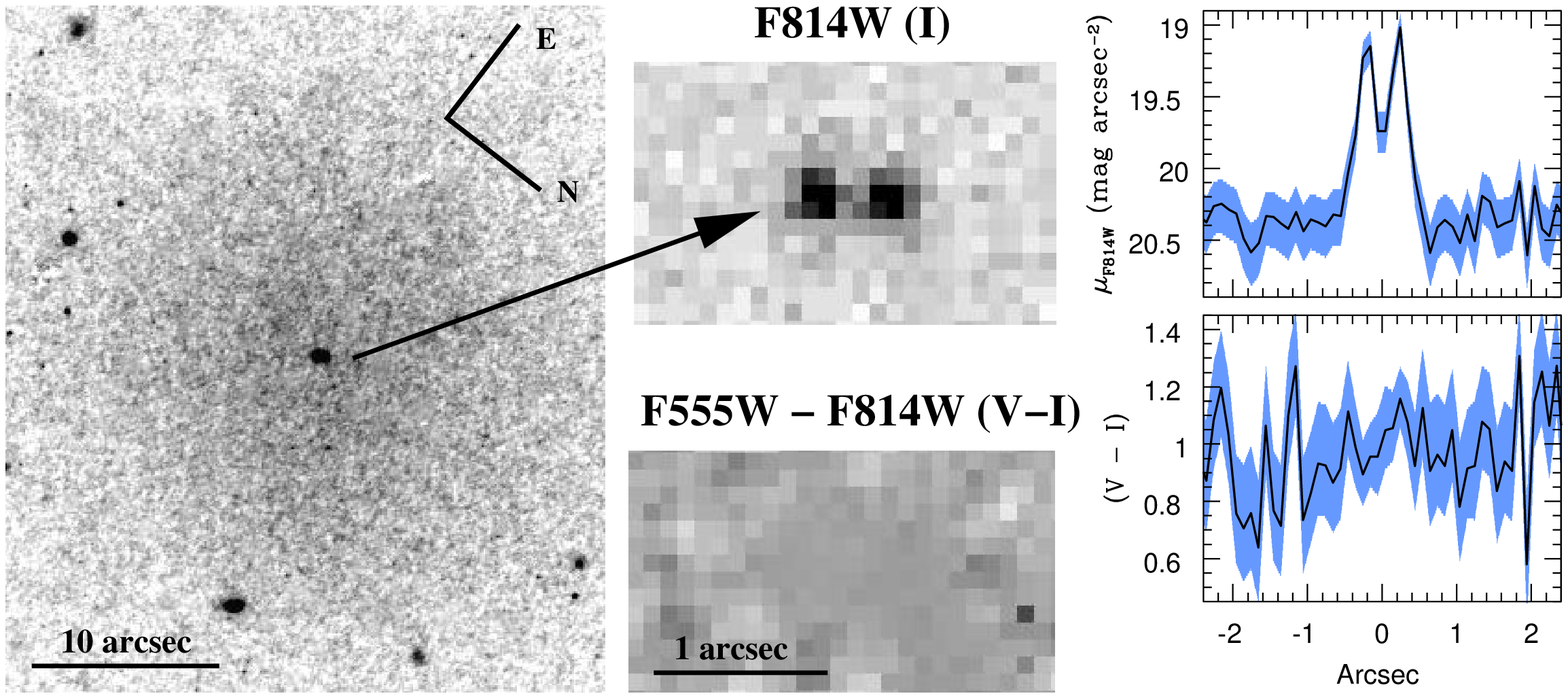}
\caption{The WFPC2 F814W image of VCC 128 (left panel, $32\arcsec
\times 37\arcsec$). In the center top panel we zoom into the central
$2\farcs 5 \times 1\farcs 5$, showing that the nucleus is resolved
into two components.  P1 is the nucleus to the south-west (left) and
P2 the one to the north-east (right).  
The nuclei are located at RA $12^h 14^m 59^s.71$ Dec $09\degrees
33\arcmin 55\farcs57$ (P1) and RA $12^h14^m59^s.73$ Dec $09\degrees
33\arcmin 55\farcs 82$ (P2) in J2000.0 coordinates.
The center bottom panel shows
the same zoom into the $V-I$ color map, where $(V-I)$ ranges between 1
and 2 mag, reaching $\sim 1.0$ in the double nucleus.  The top right
panel shows the light profile of a slit 5 pixels across along the
major-axis of the nucleus.  The shaded region indicates $\pm 1 \sigma$
error.  The bottom right panel shows the color profile for the same
slit.
\label{fig:vcc128nucleus}}
\end{figure*}

VCC 128 (UGCA 275) is a dE galaxy of $m_B = 15.6$ ($M_B = -15.5$) in
the outskirts of the Virgo Cluster.  VCC 128 was imaged for 460
seconds in the F555W ($V$-band) filter and 300 seconds in the F814W
($I$-band) filter.  Its two nuclei are separated by 4 pixels of WF3
(see Figure \ref{fig:vcc128nucleus}), \ie\ $\sim 0\farcs4$ or $32$ pc
\citep[assuming a Virgo Cluster distance of 16.5
Mpc;][]{tonry_etal_01, jer_etal_04}.  We label the component to the
south-west as P1 and the one to the north-east as P2.  These are
equally bright, with apparent magnitudes $\simeq 22.56\pm 0.05$ ($M_V
= -8.5$ in P1) and $\simeq 22.72 \pm 0.06$ (P2) in the F555W filter
and $V-I = 0.9 \pm 0.07$ (P1) and $1.15 \pm 0.07$ (P2).  The $S/N =
4.3-5.0$ in $V$ and $6.2-6.6$ in $I$ for the two nuclei measured with
an aperture of radius 2 pixels.  The $V-I$ color map (Figure
\ref{fig:vcc128nucleus}) shows that the double nucleus is not caused
by patchy obscuration.  Each nucleus is resolved, with FWHM (from a
Gaussian fit) of 3.1 and 3.9 pixels in the $I$-band, whereas the
filter point spread function FWHM is 1.3 pixels.

The surface brightness profile of VCC 128 is quite flat and ellipse
fitting in IRAF is unstable.  Using {\sc GALFIT} \citep{galfit} we
fitted the surface brightness by a S\'ersic model, after masking out
stars and the nuclei.  We obtained a S\'ersic index $n = 0.35$ and a
half light radius $R_{eff} = 9\arcsec$ for F814W and $n = 0.55$ and
$R_{eff} = 14\farcs5$ for F555W.
The circum-nuclear region (measured within $1\arcsec$ of the nuclei)
has a surface brightness $\mu_V = 21.37 \pm 0.07$ and color $\mu_V -
\mu_I \simeq 1.0 \pm 0.1$.  The average surface brightness within
$R_{eff}$ (averaged between $V$ and $I$-bands) is $\mu_V = 23.95 \pm
0.02$ mag/arcsec$^2$ and $\mu_I = 22.94 \pm 0.02$ mag/arcsec$^2$.
The midpoint between the centroids of the two nuclei (measured within
an aperture of 2 pixels) is offset by $\sim 0\farcs8$ ($\sim 60$ pc)
from the photocenter of the galaxy as fitted by {\sc GALFIT} but with
these shallow images we cannot determine whether the photocenter
varies with radius as in other offset nuclei \citep{der_deb_04}.
Possibly because of this offset, \citet{lot_etal_04} classified VCC
128 as non-nucleated.

\section{Spectral Energy Distribution}
\label{sec:sed}

\subsection{Observations and Reductions}

We obtained long-slit spectra through the nucleus of VCC 128 at the
Apache Point Observatory (APO) 3.5-m telescope using the Double
Imaging Spectrograph with $0\farcs4$ pixels.  We used the low
resolution blue grating with a resolution 2.4 \AA/pixel over the
wavelength range $3660 \leq \lambda \leq 5600$ \AA.  The data were
obtained during two runs.  In the first run on March 8, 2005 we
obtained a total of 5.3 hours of data using a $1\farcs 5$ slit.
Seeing conditions varied significantly over the night, ranging from
$0\farcs 8$ to $ > 2''$ and we had difficulty pointing at the nucleus
during some of the exposures.  Unsurprisingly, the double nucleus can
be detected only in 2 one-hour exposures from this run.  Therefore we
also obtained data during a second run on April 21, 2006 under
sub-arcsecond seeing.  During this run we used the $0\farcs 9$ slit to
obtain 2 one-hour spectra.

The spectra were reduced with IRAF using standard tasks.  Flux
calibration, using the standard stars BD182647 for the first run, and
BD262606, Feige~98 and HR~4554 for the second run, achieved a $7\%$
uncertainty at wavelengths bluer than 4000 \AA, increasing to 10$\%$
at 5500 \AA.  Data from the two runs were combined by smoothing the
2006 data to match the FWHM of the 2005 run (6\AA, or 454 $\kms$ at
the CaII lines).  The spectrum of the nucleus (spatially unresolved
from the ground) is typically 6 pixels wide along the cross-dispersed
axis, and that of the galaxy 51 pixels. We extracted the spectrum of
the nucleus from the 4 exposures, integrated over the 6 central
pixels. The galaxy's spectrum was extracted below and above the
nucleus for a total extension of 45 pixels along the spatial
direction. To remove the galaxy contamination in the spectrum of the
nucleus, we subtracted the scaled galaxy spectrum from it.  The
individual spectra of the nucleus and the galaxy were then averaged to
increase their $S/N$ ratio, which turned out to be $\sim 2$ and $\sim
8$ per pixel respectively.  The spectra of the galaxy and of the
nucleus are shown in Figure \ref{fig:c2}.  At the nucleus, $\sim 12\%$
of the light comes from the nuclei, the rest being due to the galaxy.

\subsection{Spectral analysis}

\begin{figure}[!htbp]
\plotone{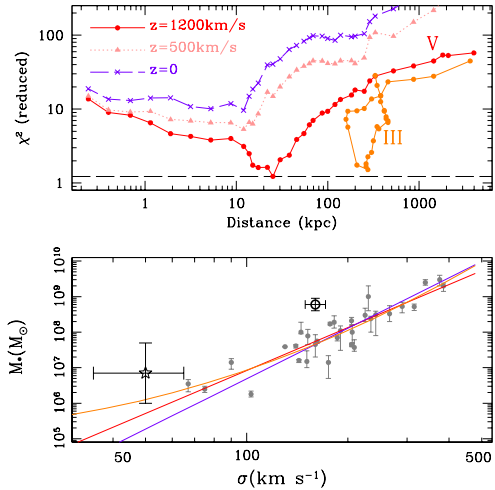}
\caption{Top: Analysis of the nuclear spectrum assuming it is a
star. The black lines compare the observed SED with a range of dwarf
stars (O-M). The gray line corresponds to a range of giant stars. We
use $V=22.6$ to derive distances.  The dashed and dotted lines
correspond to an assumed velocity $=0$ and $500 \kms$, respectively,
whereas the solid lines are computed for the best estimate --
$v=1190^{+289}_{-190} \kms$ at $95\%$ confidence level -- computed by
cross-correlating the observed SED with the stellar template that
gives the best fit (G8V, $M_V = +5.6$).  The best fit for a giant is a
G5III ($M_V=+0.4$) with $\tilde{\chi}^2=1.52$.
Bottom: The estimated location of VCC 128 (star) on the \Msig\ plane.
The solid circles are the compilation of \citet{tre_etal_02}, with the
solid line showing their best fit relation.  The dashed and dotted
lines show the relation as derived by \citet{mer_fer_01} and
\citet{wyithe_06}, respectively.  The open circle shows NGC 4486B from
\citet{kor_geb_01} and Kormendy ({\it priv. comm.}).
\label{fig:starmsig}}
\end{figure}

The absence of emission lines in the nuclear spectrum rules out, at
high confidence, that either P1 or P2 is a background emission line
object.
A foreground star can be ruled out on several counts.  The nuclei are
both resolved by {\it HST} and standard models of the Milky Way
\citep{bah_son_80} predict only 0.22 stars of this magnitude to fall
within $2 R_{eff,V} = 29\arcsec$ in this part of the sky.  We also
used our spectra to measure the velocities of the galaxy and of the
nuclei from the centroid of a Gaussian fit to the CaII H and K lines.
For the galaxy we obtained $V = 1331 \pm 184 \kms$ and for the nuclei
$1179 \pm 185 \kms$ (heliocentric).  As these lines have very low
$S/N$, we also measured velocities by modeling the entire spectral
energy distribution (SED) assuming various stellar templates from the
\citet{pickle_98} library.  We considered templates of solar
metallicity, although fits to non-solar metallicities will not change
the outcome of the analysis.
Figure~\ref{fig:starmsig} shows the reduced $\chi^2$ for $704$
degrees of freedom of a comparison in the spectral range
$3880\leq\lambda\leq 5000$\AA, which includes the 4000\AA\ break, the
G-band step, Balmer features and metallic absorption lines.
Wavelengths outside this range have very low $S/N$.  Galactic
reddening was applied using the prescription ($R=A_V/E(B-V)=3.1$) of
\citet{fitzpa_99} for a color excess of $E(B-V)=0.015$
\citep{schlegel_etal_98}.  The dots in Figure~\ref{fig:starmsig}
span the temperature range from O5 ({\sl right}) to M6 for dwarfs
(black lines) and from O8 to M9 for giants (gray line).
The $\tilde{\chi}^2$ is shown as a function of distance, which is
computed by comparing the apparent and absolute magnitudes
corresponding to each star.  Overall, the best fitting foreground star
would be a G8V dwarf at $25$ kpc ($\tilde{\chi}^2=1.23$) with
$V=1190^{+289}_{-190} \kms$.  All fits with $\tilde{\chi}^2 < 2$
require $v > 500 \kms$, which is improbable for stars in the Milky Way
halo.  Hence, neither source at the center of VCC 128 can be a
foreground star but in good agreement with Virgo Cluster membership,
leading us to conclude that this is the nucleus of VCC 128.

\begin{figure}[!htbp]
\plotone{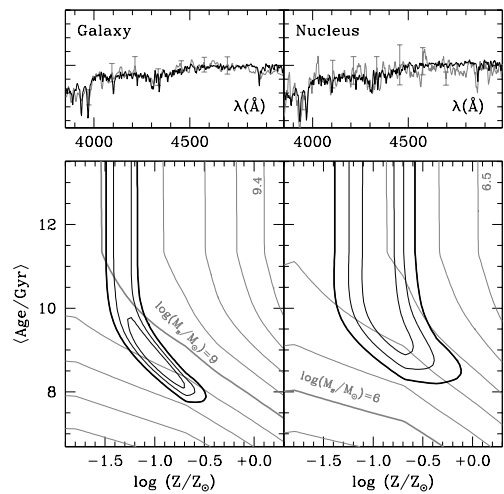}
\caption{Top: the observed (smoothed) SEDs.  The two panels compare the
SED of the galaxy (left) and the nucleus (right) (gray line), with the
best fitting model SED (black lines).  The feature at $\sim 4800$ \AA\
in the observed spectra is due to an artifact of the flux calibration
due to a 'bump' in the grating response, while that at $\sim 4600$
\AA\ in the nuclear spectrum is a remnant of the background
subtraction.
Bottom: the corresponding mean properties of the best-fitting stellar
populations.  Confidence levels (black lines) are given at the $1$,
$2$ and $3\sigma$ (thick lines) levels.  Contours of stellar mass are
shown in both cases (gray lines).  For the nucleus, the mass shown is
for the two components combined.
\label{fig:c2}}
\end{figure}

To estimate the mass of the nucleus, we compared the SEDs of both the
galaxy and the nucleus with the population synthesis models of
\citet{bru_cha_03}.  We generated a $64\times 64$ grid of
$\tau$-models, \ie\ composite stellar populations according to an
exponentially decaying star formation rate that are then modeled at
the resolution and pixel sampling of the observations, and compared
via a maximum likelihood analysis.  The formation epoch was fixed at
$z_F=3$. The two parameters explored in the grid are the exponential
star formation timescale ($-1<\log (\tau /{\rm Gyr})<1$) and the
metallicity ($-2<\log (Z/Z_\odot )<+0.3$).
Figure~\ref{fig:c2} shows the $1$, $2$ and $3\sigma$ (thick lines)
confidence levels for the analysis of the galaxy ({\sl left}) and of
the nucleus ({\sl right}).  The plots overlay contours of stellar mass
corresponding to the observed apparent magnitude assuming a
\citet{chabri_03} initial mass function (IMF), and using the total
apparent magnitude of the galaxy \citep[$m_B=15.6$][]{bing_etal_85}
and for the double nucleus combined.  Our results for the ages and
metallicities are independent of the assumed IMF but the masses are
more sensitive to this change.
The nucleus is consistent with having a relatively metal rich stellar
population older than 8 Gyr.  With the assumed IMF, the best-fit
combined nuclear stellar mass is $\sim 10^6\Msun$.  The galaxy
spectrum gives -- within error bars -- a similar value for the age and
metallicity.

In order to assess the effect of dust on the SED analysis, we ran a
few sets of grids for various values of $E(B-V)$.  The best fits
obtained for these realizations gave consistently higher values of
$\chi^2$, so that a color excess above $E(B-V)>0.1$~mag. is ruled out
at the $3\sigma$ level.

\section{Discussion and Conclusions}
\label{sec:discussion}

Since the nucleus is old, the present configuration could not have
resulted from recent gas infall.  It is unlikely, though not
impossible, that the two nuclei are both globular clusters 32 pc apart
in the last stages of merging because of the short lifetime of such a
configuration ($\sim 5$ Myr using the standard \citet{chandr_43}
formula with $\ln \Lambda = 4$ and $v_c = 20 \kms$).  Two
well-separated globular clusters projected so close to each other, so
close to the center of the galaxy, and so similar in their properties
is even more unlikely.
The double nucleus in VCC 128 is similar in many ways to that in NGC
4486B.  As in NGC 4486B, a nuclear disk surrounding a SMBH provides
the best explanation: this would be a stable, long-lived
configuration, and would account for the similar colors and magnitudes
of the two components and for their location near the center of the
galaxy.

If the double nucleus is a disk orbiting a SMBH, we can estimate a
lower limit to \Mbh\ by assuming that it is larger than the nuclear
disk mass, ${\rm M}_d = 10^6 \Msun$.
Alternatively we can use the ratio of ${\rm M}_d$, to \Mbh\ in M31
\citep[${\rm M}_d/\Mbh =0.16$][]{tremai_95} and in NGC 4486B
\citep[${\rm M}_d/\Mbh = 0.019$][]{kor_etal_97, lauer_etal_96} to
estimate $\Mbh \sim 6 \times 10^6 - 5 \times 10^7~ \Msun$ in VCC 128.
We estimate a velocity dispersion, $\sigma$, for VCC 128 from the
Faber-Jackson relation of dE's \citep{der_etal_05}.  We obtain $\sigma
\sim 35-65 \kms$.  Thus within large uncertainties the postulated SMBH
in VCC 128 could satisfy the \Msig\ relation, as we show in Figure
\ref{fig:starmsig}.

\citet{ferrar_02} proposed that galaxies with a dark matter virial
velocity $v_{vir} \la 200 \kms$ would not be able to form SMBH's.
Using her scaling relations between $\sigma$ and $v_{vir}$, we find
that $v_{vir} \la 65 \kms$ assuming the $\sigma$ estimated above.  It
is unlikely that there is enough scatter in the scaling relations as
to accommodate $v_{vir} \simeq 200 \kms$ in VCC 128.  Thus its SMBH
probably violates the proposed limit unless VCC 128 has been heavily
stripped in the cluster environment.  Curiously however, VCC 128 would
still satisfy the relation between $M_{\rm CMO}$ and $B-$band
magnitude $M_B$ \citep{weh_har_06, fer_etal_06}.

We have shown that the nucleus of VCC 128 is double, with the two
components of equal magnitude and color.  We proposed that this can be
explained by the presence of a disk surrounding a SMBH.  Simple
estimates of its mass all put it in the regime $\Mbh \sim 10^6~\Msun$
or larger.  If VCC 128 is on the fundamental plane, then these
estimates put the SMBH at, or above, the \Msig\ relation.  However the
halo virial velocity would be smaller than the proposed limit for
galaxies to be able to form SMBHs.  Amongst the dwarf
elliptical/spheroidal population of galaxies, this is only the second
example of a system with evidence for a SMBH
\citep{maccarone_etal_05}.  Given the very interesting nature of this
object, we suggest that further high-resolution imaging and
spectroscopy of VCC128 would be well worthwhile.


\acknowledgments VPD is supported by a Brooks Prize Fellowship at the
University of Washington and receives partial support from NSF ITR
grant PHY-0205413.  SDR acknowledges Postdoctoral Fellowship support
from the Fund for Scientific Research--Flanders, Belgium (FWO).  VPD
thanks the MPIA Heidelberg for hospitality during part of this
project.  We thank the anonymous referee for comments that helped to
significantly improve this paper.

\bibliographystyle{aj.bst}
\bibliography{allrefs}


\end{document}